# Deep Knowledge Tracing is an implicit dynamic multidimensional item response theory model

**Jill-Jênn VIE[a*], Hisashi KASHIMA[b]**
[a]*Soda Team, Inria, France*
[b]*Kyoto University, Japan*
*jill-jenn.vie@inria.fr

**Abstract:** Knowledge tracing consists in predicting the performance of some students on new questions given their performance on previous questions, and can be a prior step to optimizing assessment and learning. Deep knowledge tracing (DKT) is a competitive model for knowledge tracing relying on recurrent neural networks, even if some simpler models may match its performance. However, little is known about why DKT works so well. In this paper, we frame deep knowledge tracing as a encoder-decoder architecture. This viewpoint not only allows us to propose better models in terms of performance, simplicity or expressivity but also opens up promising avenues for future research directions. In particular, we show on several small and large datasets that a simpler decoder, with possibly fewer parameters than the one used by DKT, can predict student performance better.

## 1. Introduction

Adaptive testing and personalized learning are precious technologies that have been enabled by tracing the knowledge of previous students. They rely on a response model of the learners: if we know how people learned in the past, we can optimize testing or learning for new students. This is the model-based approach of reinforcement learning, commonly referred to as knowledge tracing in the educational data mining literature.

Formally, knowledge tracing relies in predicting the outcomes of students on some items, given their previous interactions on some items. Numerous models have been proposed for knowledge tracing such as deep knowledge tracing (DKT) proposed by Piech et al. (2015) that relies on a recurrent neural network (RNN). However, Wilson et al. (2016) have matched the performance of DKT with a unidimensional item response theory (IRT) model that can be seen as online logistic regression. In order to advance the field, we need to understand what distinguishes these models, and how to build upon them. For example, we can notice that these models do not rely on the exact same data. IRT usually learns a difficulty parameter per item while DKT were initially strictly used on skill data for the sake of the comparison with Bayesian knowledge tracing (BKT).

In this paper, we show how we can bridge both categories of models using an encoder-decoder architecture. These architectures are usually encountered in sequence-to-sequence scenarios such as machine translation (Cho et al., 2014) and got popular with the rise of transformers (Vaswani et al., 2017). We open the "black box" of DKT and explain how it relates to existing, well known models.

Our main contribution is an encoder-decoder architecture that takes DKT, IRT and other models such as Performance Factor Analysis (PFA) as special cases. We demonstrate using several real datasets that it seems better to learn unidimensional parameters for item, skills, which is not what the vanilla DKT model does, relying instead on multidimensional parameters for skills. In this paper, we are not so much interested in state-of-the-art performance but in understanding how models are related to each other and how to interpret their parameters, so that they can also be visualized. We notably show that metadata considered can have a considerable impact on performance.

We first expose related work, then our approach: summarize & predict. We subsequently detail our datasets, experiments, and see the influence of specific components of our architecture in the results.

## 2. Background and related work

Formally, knowledge tracing can be defined as follows. Let us denote $I$ the set of items of some test, and $K$ the number of knowledge components assessed. For each student, at each time step $t$, we know the sequence of items and outcomes already given to the student $(q_{1:t}, a_{1:t}) = (q_\tau, a_\tau)_{1 \leq \tau < t}$ (here, time $t$ is excluded as it is what we want to predict) where $q_\tau \in I$ and $a_\tau \in \{0,1\}$. We may as well have some side information about $q_t$ such as the knowledge components (KCs) required by $q_t$ denoted $KC(q_t) \subset \mathcal{P}(\{1, \ldots, K\})$, sometimes represented by a binary vector from $\{0,1\}^K$ in a so-called $Q$-matrix that indicates which knowledge components are assessed. Using this information, we need to predict outcome $a_t$ using a model that estimates a probability $p_t$. We will note $\sigma$ the sigmoid function : $\sigma: x \mapsto 1/(1+\exp(-x))$. We now describe some models for knowledge tracing, of two different kinds.

### 2.1. Sequence-based architectures

**Deep knowledge tracing models.** Deep knowledge tracing (DKT) is usually described as a black-box model that takes into input pairs $(q_{1:t}, a_{1:t}) = (q_\tau, a_\tau)_{1 \leq \tau < t}$ from some student, and outputs a vector of probabilities $y$ such that $y_k$ is the probability that the student will answer correctly an item requiring skill $k$. More precisely, in a DKT model, the actual probability that a certain user correctly answers an item that requires skill $k$ at time $t$ (event $R_{tk}=1$) is given by:

$$Pr(R_{tk} = 1) = \sigma(\langle h_t, v_k \rangle + w_k)$$

where $v_k \in \mathbf{R}^d, w_k \in \mathbf{R}$ is a representation of the skill $k$ learned by DKT that does not evolve over time and $h_t = LSTM(h_{t-1}, q_t, a_t)$ if $t > 1$, $h_1 = 0$ otherwise, is a representation of the user that evolves over time. Montero et al. (2018) have shown that DKT can trace knowledge efficiently even when skills are interleaved, and that it shares information between skills. In the original DKT paper, Piech et al. (2015) assume that each item is only related to one skill among $1, \ldots, K$. They learn a joint representation in $\{0,1\}^{2K}$ for the pair $(s_t, a_t)$ where $s_t = KC(q_t)$, because they claim that separate representations for $s_t$ and $a_t$ degraded performance. When there are too many pairs $(s_t, a_t)$, they use fixed low-dimensional representations instead of $\{0,1\}^{2K}$, which can be seen as setting a random embedding for each pair. It makes sense to use skill instead of items for encoding the actions of the student, in order to have redundancy and avoid the item cold-start problem. The drawback is that in their vanilla form, DKTs cannot handle multiple skills per item. In this paper, we show in particular how multiple skills can be taken into account, which can be used to improve DKT.

In DKT-DSC, an extension of DKT, Minn et al. (2018) use an encoding of the triplet $(s_t, a_t, c_t)$ as input metadata where $c_t$ is a dynamic clustering information of the student based on a vector counting successful and unsuccesful attempts, updated at each time step. They managed to outperform DKT on several datasets.

Ding & Larson (2019) have shown that the encoder of DKT with untrained, random weights, is a strong baseline; which seems to indicate that the latent state $h_t$ can be a random embedding of the sequence while the final fully connected layer is mainly responsible for the good predictive performance.

More recently, transformers have been proposed for knowledge tracing (Pandey & Karypis, 2019; Ghosh, Lan, & Heffernan, 2020). Self-attention can be used to model how the samples we want to predict relate to previous observations from the student.

*2.2. Non-sequence-based architectures*

**Multidimensional item response theory (MIRT).** In item response theory, one usually does not assume that the examinee's ability evolves over time. The probability that user $i$ correctly answers item $j$ (event $S_{ij} = 1$) is:
$$\Pr(S_{ij} = 1) = \sigma(\langle \theta_i, v_j \rangle + w_j)$$
where $\theta_i \in \mathbf{R}^d$ is a learned representation of user $i$ called multidimensional ability, $v_j \in \mathbf{R}^d$ is a learned representation of item $j$ called discrimination parameter and $w_j \in \mathbf{R}$ is a bias parameter representing the easiness of item $j$. This model is popular in the psychometric literature, because it can enable computerized adaptive testing: adapting the assessment, given the currently estimated performance of the student. It is indeed possible to make the model dynamic by updating the maximum likelihood estimate (MLE) of $\theta_i$ at each new observed point, which can take time to estimate. This is exactly what Wilson et al. (2016) do, to demonstrate a better performance of updated IRT compared to DKT. Maximum likelihood estimation (MLE) is possible with simple models such as IRT, but for more complex architectures or priors, it is not always feasible because it increases time complexity during the test phase. This is why simpler online updates have been proposed such as multivariate Elo (Abdi et al., 2019) to train dynamic MIRT models.

**Performance Factor Analysis and count-based models.** Several models rely on attempt counts, such as Additive Factor Model (AFM) and Performance Factor Analysis (PFA) proposed by Pavlik, Cen, & Koedinger (2009). The estimated ability of the student at time $t$ is a linear function of their number of previous successful $W_k^{1:t}$ and unsuccessful attempts $F_k^{1:t}$ over skill $k$ up to time $t$ (again, time $t$ information is excluded, as it is not observed yet at prediction time). An important remark is that here, we forget the order in which the items were solved, we are considering mere counts. DAS3H proposed by Choffin et al. (2019) maintains counters in several time windows, which enables the modeling of forgetting. In PFA, $p_t$ depends on a weighted sum of the counts, according to the knowledge components that are assessed at time $t$ and specified in $KC(q_t)$:
$$D_{\text{PFA}} : p_t = \sum_{k \in KC(q_t)} \beta_k + \gamma_k W_k^{1:t} + \delta_k F_k^{1:t}$$
where parameters $w = (\beta, \gamma, \delta)$ are learned, respectively scalar parameters for skills, wins, fails. Please note that this formulation can handle multiple skills.

**Knowledge Tracing Machines.** Vie & Kashima (2019) have shown that it is possible to learn representations for all users, items, skills in a test and combine them in a pairwise manner. Interestingly, most existing models for knowledge tracing such as IRT, PFA, MIRT are special cases of KTMs, according to the representations considered in the modeling. In this paper we want to emphasize the importance of metadata. Gervet et al. (2020) have also shown that predicting performance at the item level instead of at the skill level could improve performance of DKT.

## 3. Our approach: summarize & predict

We see all these models through the same lens: our architecture relies on two main components, an encoder $E$ that summarizes prior data and a decoder $D$ that predicts student performance; both are trained jointly. They respectively require an expression of the metadata of former actions from the student $(q_{1:t}, a_{1:t})$ and metadata of assessment at time $t$: which item is being attempted, which skills are being assessed, etc.
$$\begin{cases} h_t = E((q_{1:t}, a_{1:t})) \\ p_t = D(h_t, q_t) \end{cases} \quad t = 1, \ldots, T$$

In particular, $h_1 = E(\varnothing) = 0$ where $\varnothing$ denotes the empty sequence. In this expression, $h_t$ represents the learned representation of the user at time $t$, and $p_t$ the probability that the attempt of the student at time $t$ will be correct. Please note that this modeling is rich enough to encompass broader tasks than knowledge tracing, for example we can model the event that some user watched the video of some lesson, with the corresponding metadata. If we can take advantage of this kind of data, we can hopefully recommend new videos to some student to optimize their learning.

## 3.1 Summarize: encoding a dynamic representation of what happened

The encoder $E$ is a function of past history. It can either be simple counts to remember how many times a student succeeded or failed an attempt at a single skill, or a more complex function like a RNN in the case of DKT: $E_{DKT}: h_t = RNN(h_{t-1}, q_t, a_t)$ if $t > 1$, $h_1 = 0$ otherwise. In order to share information across items, it is common to use pairs $(s_t, a_t)$ as sequence inputs where $KC(q_t) = \{s_t\}$. If $q_t$ is associated with multiple KCs, some tricks consist in taking one at random, or creating a new skill that represents the combination of multiple skills.

## 3.2 Predict: combine the learned representations to make a prediction

The objective of the decoder $D$ is to combine the learned representation $h_t$ of the student at time $t$ with some parameters involved in the assessment at time $t$ such as which item was administered, which skills were assessed, etc.

In order to build a decoder we need to decide two main characteristics: what is the metadata taken into account for the prediction (item $q_t$, skills $KC(q_t)$, wins $W_k^{1:t}$, fails $F_k^{1:t}$), and what is the embedding size of those features.

### 3.2.1 Output metadata

If the output is solely based on item ID, we get a model like IRT or MIRT according to the embedding size of items. If the output is based on skill IDs $KC(q_t)$, number of successful attempts (wins $W_k^{1:t}$), number of unsuccessful attempts (fails $F_k^{1:t}$), and the embedding size is 1, we get a model similar to PFA.

### 3.2.2 Decoder of same embedding size as $h_t$

As an example, MIRT computes a dot product of the user embedding (usually denoted as $\theta_t$ in the literature, here $h_t$ for consistency) and the embedding of item $q_t$ (usually called discrimination, here $v_{q_t}$), plus a bias representing the easiness of the item (here, $w_{q_t}$):
$$D_{\text{MIRT}} : p_t = \sigma(\langle h_t, v_{q_t} \rangle + w_{q_t})$$
The vanilla DKT model has a fully connected layer that converts the latent state $h_t$ to the prediction of performance of every possible skill. In practice, we are only interested in the skill $\{s_t\} = KC(q_t)$ assessed at time $t$ (assumed single by DKT) and DKT implicitly computes a dot product of the user embedding $h_t$ and the skill embedding $v_{s_t}$, leading to the following expression:
$$D_{\text{DKT}} : p_t = \sigma(\langle h_t, v_{s_t} \rangle + w_{s_t}).$$
Now, what should strike the reader is that the expressions of $D_{\text{MIRT}}$ and $D_{\text{DKT}}$ are almost the same: only the metadata is different, item ID for MIRT and skill ID for DKT. This is why DKT, in its item output form, can be seen as a dynamic MIRT model, where a RNN looks at the sequence of observations to estimate ability parameters at current time. See Figure 1 for an illustration.

In fact, Gervet et al. (2020) have shown an improvement in performance when considering items as output of DKT instead of skills as output.

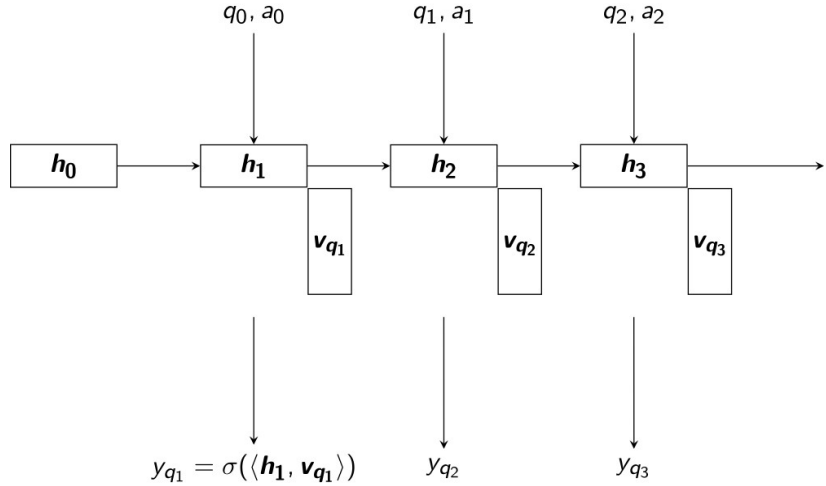

*Figure 1.* DKT is implicitly a dynamic MIRT model.

*3.2.3 Decoder of different embedding size than encoder*

The encoder usually needs a high dimensional embedding to encode learning dynamics, notably if it is a RNN. The item and skill embeddings do not need to be the same dimension as the latent state $h_t$. Indeed, in practice, most hard baselines for knowledge tracing such as PFA only model skill parameters with one dimension. This is why we suggest reducing the dimension of $h_t \in \mathbf{R}^d$ before making the predictions, using a fully connected layer $d \to d'$:

$$h'_t = A h_t + b$$

where parameters $A \in \mathbf{R}^{d' \times d}$ and $b \in \mathbf{R}^{d'}$ are learned. In the most extreme example, $h'_t$ can be scalar, i.e. unidimensional, $d' = 1$. The advantage is that it drastically reduces the number of parameters. For example, a decoder of embedding size 1 using (item, skill, wins, fails) as metadata would be a blend of IRT and PFA models and would look like this:

$$D_{\text{iswf}} : p_t = \sigma \left( h'_t + \underbrace{w_{q_t}}_{\text{item}} + \sum_{k \in KC(q_t)} \underbrace{\beta_k}_{\text{skill}} + \underbrace{\gamma_k W_k^{1:t}}_{wins} + \underbrace{\delta_k F_k^{1:t}}_{\text{fails}} \right)$$

where weights $w$ are learned for each item and weights $\beta, \gamma, \delta$ are learned for each skill. We will denote such a decoder as "iswf $d' = 1$" later on. Another advantage of this decoder compared to vanilla DKT is that it can consider multiple skills associated to $q_t$. To see how embeddings of higher dimension $d' > 1$ can be considered for multiple metadata, see Vie & Kashima (2019).

*3.3 Evaluation*

We optimize the log-loss, also known as mean negative log-likelikood:

$$L(a,p) = \sum_{t=1}^{T} \log(1 - |a_t - p_t|).$$

This log-loss can be computed for a batch of students, but we have to take into account that all students did not attempt the same number of items, so student sequences within a batch may have different lengths.

Once the parameters of the encoder $E$ and the decoder $D$ have been trained, for a new student it is easy to unroll the encoder on the sequence of outcomes $(q_t, a_t)$ encoded as input metadata, get the corresponding latent states $h_t$, and feed them to the decoder with the output metadata to get the predictions, and compute performance metrics.

We report all results according to their accuracy (ACC) and area under the curve (AUC).

## 4. Experiments

The code of our experiments is on GitHub: https://github.com/jilljenn/dktm.

*4.1 Models*

We tried different combinations of encoders and decoders. As sequence-based encoder, we used a RNN of latent dimensionality $d$ where the encoding of actions $(q_t, a_t)$ is sampled from a Gaussian multivariate distribution of dimension $d$ and fixed. We used GRU (Gated Recurrent Unit) because it has fewer parameters than LSTM (Long Short-Term Memory), so it is less prone to overfitting (Chung et al., 2015). In the reported results, when encoder is listed as "none", it means only the wins and fails parameters $W_k^{1:t}$ and $F_k^{1:t}$ were counted. Decoders are described by a $n$-gram among letters "iswf", respectively item, skill, wins and fails; and an embedding size $d'$. Examples of decoders are given in the previous section. The corresponding baselines are:
- DKT, which has a RNN as encoder and multidimensional embeddings within its decoder. The decoder can either be "s" if DKT makes predictions at skill level or "i" if DKT makes predictions at item level.
- PFA just keeps track of successful and unsuccessful attempts so there is no sequence-based encoder; important metadata is skill, wins, fails so "swf" and the embedding size is 1 as scalar parameters are learned.
- Decoders "swf $d' = 1$" (PFA) and "iswf $d' = 1$" are particular cases of logistic regression (LR) with sparse features.

*4.2 Datasets*

We tried our approach on the following datasets that have diverse characteristics.

**Fraction.** 536 middle-school students attempting 20 fraction subtraction exercises requiring 8 KCs such as, being able to put fractions at the same denominator. All students attempted all items, so this dataset is fully specified. The dataset and description of KCs can be seen in (DeCarlo, 2010). This dataset is particularly interesting because it is small scale, so complex models may overfit. Also, most items measure several KCs at the same time. As we have access to the full response pattern of all students, we assume that items are solved from item 1 to item 20.

**Assistments 2009.** 346,860 attempts of 4,217 students over 26,688 math items requiring 123 KCs (Feng et al., 2019). Some of the items require multiple KCs, up to 4. Students attempted between 1 and 1,382 exercises.

**Berkeley.** 562,201 attempts of 1,730 students over 234 CS-related items in a MOOC provided by Berkeley. Items can be of 29 categories, that we used as KCs. Each item belongs to a single category. Students attempted between 2 and 1,020 exercises. This dataset is private.

*4.3 Data preparation*

We used 5-fold cross-validation: we split each dataset into 5 folds, predict any of them using the remaining ones, and average the results. In practice there are many challenges, fortunately already known in the natural language processing community.

**Batching.** Whenever there are too many data points, in order to save memory we need to sample a batch of users. So the log-loss is computed on a batch.

**Sequences of uneven lengths.** Within a batch, different users may have attempted a different number of questions. However for efficiency, it is better to vectorize as much as possible and compute the log-loss over matrices. So we use a mask to know where the sequences end for each student.

**Long sequences.** On long sequences, DKT is hard to train. First, it may forget information on a long sequence, second the computation of the gradient may take time, vanish to zero or conversely get arbitrary high. As a remedy, we train on windows of fixed size called BPTT: backpropagating through time (Werbos et al., 1990). The latent state of the previous batch is fed to the next batch.

*4.4 Implementation details*

We optimize our model using the Adam optimizer (Kingma & Ba, 2014) with learning rate $\gamma=0.005$ and weight decay $\lambda=0.0005$ (equivalent to $L_2$ regularization). We train using 100 minibatches and use time windows of size 100. For the RNN, we used GRU with one layer, no dropout and latent dimensionality 2 or 50. Our implementation is in PyTorch. We ran the Fraction dataset experiments on CPU and the Assistments and Berkeley dataset experiments on GPU. Training was stopped after 200 epochs.

## 5. Results and Discussion

Results are shown in Tables 1 to 3 where the best models are shown in bold.

Table 1. *Results on the Fraction Dataset*

| Model | Encoder | Decoder | ACC | AUC |
|---|---|---|---|---|
| **Ours** | GRU $d=2$ | iswf $d'=1$ | **0.880** | **0.944** |
| LR | None | iswf $d'=1$ | 0.853 | 0.918 |
| PFA | None | swf $d'=1$ | 0.854 | 0.917 |
| DKT | GRU $d=2$ | i $d'=2$ | 0.772 | 0.844 |
| DKT | GRU $d=2$ | s $d'=1$ | 0.761 | 0.839 |

Table 2. *Results on the Assistments Dataset*

| Model | Encoder | Decoder | ACC | AUC |
|---|---|---|---|---|
| **LR** | None | iswf $d'=1$ | **0.714** | **0.748** |
| Ours | GRU $d=50$ | iswf $d'=1$ | 0.711 | 0.726 |
| DKT | GRU $d=50$ | i $d'=50$ | 0.691 | 0.701 |
| PFA | None | swf $d'=1$ | 0.682 | 0.686 |

Table 3. *Results on the Berkeley Dataset*

| Model | Encoder | Decoder | ACC | AUC |
|---|---|---|---|---|
| **Ours** | GRU $d=50$ | iswf $d'=1$ | **0.707** | **0.778** |
| **LR** | None | iswf $d'=1$ | **0.704** | **0.775** |
| DKT | GRU $d=50$ | i $d'=50$ | 0.684 | 0.751 |
| PFA | None | swf $d'=1$ | 0.630 | 0.683 |

For all datasets, we discovered models that outperformed the vanilla DKT model. More precisely, decoders of embedding size $d'=1$ outperform decoders of higher embedding size. This seems to indicate that, although sequence-based encoders (GRU) may be better in some cases to learn the dynamics of student knowledge, it is always better to represent assessment information (items, skills) as scalars, to avoid overfitting. Besides improving the performance, reducing parameters induces a speedup in training and allows interpretability.

On the large datasets Assistments and Berkeley, logistic regression, that uses a simple counter of successful and unsuccessful attempts at skill level, is among the top models. It seems to indicate that for big datasets, logistic regression may be enough, because the number of prior successes and failures is enough to predict. What is even more surprising, is that on the small dataset Fraction, the top performing model uses a recurrent neural network as encoder. It may be because the sequences are small: every student only attempts 20 items. So the sequences of successes and failures make students more unique than their counters of successes, inducing a more personalized estimation.

Considering an item bias does not improve a lot the quality of the predictions on the Fraction dataset, maybe because the fraction subtraction task is particularly easy to describe using KCs, so the KCs are enough to characterize the items. However, on Assistments, one can see the improvement when considering an item bias, see the difference between PFA and LR (0.06 improvement in AUC). Same goes for Berkeley (0.09 improvement in AUC), where the KCs are actually mere categories of items.

## 6. Conclusion

Most models in the educational data mining literature model users, items and skills with multidimensional parameters. In this study though, we showed that while it is indeed important to model the dynamics of the students with multidimensional parameters, the items do not necessarily need to be multidimensional to come up with really strong models. In particular, logistic regression is among the top models. This is encouraging because we come up with better models that have fewer parameters. DKT may be slow to train not because of the recurrent neural network but because of the fully connected layer at the output, that contains the most parameters. In this paper, we suggest to replace it with a lower-dimensional layer.

As future work, we plan to try to use other kinds of side information. For example, for the Duolingo SLAM dataset (Settles et al., 2018) for second language acquisition modeling, contestants had to predict whether a learner would get a word correct, and the best approaches were combinations of DKT with word embeddings (Mikolov et al., 2013). Using word content, be it only $n$-gram features, allows to learn what phonemes are harder for which categories of people. Also we will try to encode richer actions for the encoder, because it is costly for humans to define the knowledge components related to all items, and deep learning models may be able to recover most of those with little supervision, reducing the workload on practitioners.

## 7. Acknowledgements

We thank Samuel Girard for his insightful comments. This work has been supported by JST CREST Grant Number JPMJCR21D1.